\documentclass[referee, useAMS, usenatbib]{mn2e}  
\usepackage{psfig}
\usepackage{graphicx}
\usepackage{epsf}
\usepackage{bm} 
\title [Dynamical evolution of active detached binaries and contact binary formation]
{Dynamical evolution of active detached binaries on $\log J_{o}-\log M$ diagram and contact binary formation}

\author[Eker et al.]
       {Z. Eker,$^1 \thanks{E-mail: eker@comu.edu.tr}$
        O. Demircan$^{1}$, S. Bilir$^{2}$ and Y.~Karata\c{s}$^2$ \\ 
  $^1$\c{C}anakkale University Observatory, 17100 \c{C}anakkale, Turkey\\
  $^2$Istanbul University Science Faculty, Department of Astronomy and Space
      Sciences, 34119, University-Istanbul, Turkey\\}

\date{Accepted 2006 month day.
      Received year month day;
      }

\pagerange{\pageref{firstpage}--\pageref{lastpage}}
\pubyear{2006}

\begin{document}

\maketitle

\label{firstpage}

\begin{abstract}

Orbital angular momentum ($J_{o}$), systemic mass ($M$) and orbital period ($P$) distributions of chromospherically active binaries (CAB) and W Ursae Majoris (W UMa) systems were investigated. The diagrams of $\log J_{o}-\log P$, $\log M-\log P$ and $\log J_{o}-\log M$ were formed from 119 CAB and 102 W UMa stars. The $\log J_{o}–-\log M$ diagram is found to be most meaningful in demonstrating dynamical evolution  of binary star orbits. A slightly curved borderline (contact border) separating the detached and the contact systems was discovered on the $\log J_{o}–-\log M$ diagram. Since orbital size ($a$) and period ($P$) of binaries are determined by their current $J_{o}$, $M$ and mass ratio $q$, the rates of orbital angular momentum loss ($d\log J_{o} /dt$) and mass loss ($d\log M/dt$) are primary parameters to determine the direction and the speed of the dynamical evolution. A detached system becomes a contact system if its own dynamical evolution enables it to pass the contact border on the $\log J_{o}-\log M$ diagram. Evolution of $q$ for a mass loosing detached system is unknown unless mass loss rate for each component is known. Assuming $q$ is constant in the first approximation and using the mean decreasing rates of $J_{o}$ and $M$ from the kinematical ages of CAB stars, it has been predicted that 11, 23 and 39 cent of current CAB stars would transform to W UMa systems if their nuclear evolution permits them to live 2, 4 and 6 Gyrs respectively.     

\end{abstract}

\begin{keywords}
stars: mass-loss, stars: evolution, stars: binaries spectroscopic 

\end{keywords}

\section{Introduction}
The angular momentum loss in close binaries is known to be replenished from the reservoir of orbital angular momentum (OAM) of the system by tidal locking. Since tidal locking operates as a mechanism to draw angular momentum from the orbit, the orbits of spin-orbit coupled binaries are then forced to shrink although actual angular momentum loss occurs at one or both components due to magnetically driven winds, also called magnetic braking \citep{Sch59,K67, M68}. Shrinking orbits, then, require orbital periods to be decreasing. Authors such, \cite{Huang66, OS70, van79, VR80, R82, M84, GB88, MV91,  MV96, S95, D99} all believed this mechanism is a main route to form W UMa  systems from the binaries initially detached with comparable periods. Showing high level of chromospheric and magnetic activity, RS CVn like objects including binary BY Dra type systems, shortly called CAB, are primary candidates to be the progenitors of the contact systems. 

\cite{K04} recently presented observational evidences of mass loss and orbital period decrease by comparing the total mass and period histograms between the kinematically young (age 0.95 Gyr) and older (age 3.86 Gyr) CAB sub samples. Secular orbital period decrease among CAB is indicated further by the fact that the shorter orbital period systems are older than the longer orbital period systems \citep[see Table 5 of ][]{K04}. Moreover, 5.47 Gyr kinematical age for the field W UMa stars determined by \cite{B05} appears consistent with the scenario. Consequently, the mean difference of 1.61–-Gyr age between field W UMa and field CAB could be interpreted as a mean lifetime of contact stages. Not all contact systems have to be formed from detached progenitors. This is because \cite{B05} have found that about 20 per cent of contact binary sample have ages less than 0.6 Gyr. Such a young age, being much less than the mean life time (1.61--Gyr) of contact stages, does not permit pre-contact detached phases. A Pre--contact detached stage usually has a duration up to several Gyr, e.g. 3.86 Gyr is an average. Therefore, those young W UMa must have been formed directly in the beginning of the main sequence or during the pre--main sequence contraction phase. Apparently, both mechanisms with and without pre-contact phases actively operate.

Recently, \cite{D06} have determined a first order relative decreasing rates of OAM, systemic mass and orbital period for the detached CAB as $\dot J_{o}/J_{o}=-3.48\times 10^{-10}~yr^{-1}$, $\dot M/M=-1.30 \times 10^{-10}~yr^{-1}$ and $\dot P/P=-3.96\times 10^{-10}~yr^{-1}$ by using kinematical ages of sub samples formed according to OAM ranges. Mass loss, OAM loss and orbital period decrease driving one another appear to be confirmed by observational data. Nevertheless, it has been known that an isotropic mass loss from the surface(s) of one/or both component(s), despite it means OAM loss, forces the orbit to be bigger and makes the orbital period larger \citep{P85}. It has also been shown by \cite{D06} that not all rates of OAM loss would cause an orbit to shrink. Only if there is an amplification mechanism, such as magnetic loops forcing magnetized plasma to co--rotate, so mass loss at the Alfen radius carries away more angular momentum. If the amplification parameter, $\bar A=(dJ_{o}/dM)/(J_{o}/M)$, is bigger than 5/3, then  the binary orbit would shrink and the orbital period would decrease. With the decreasing parameters given above, \cite{D06} found $\bar A=2.68$ as a first order observational mean.

Although, shrinking orbits and decreasing orbital periods are confirmed observationally and mean decreasing rates are available to support dynamical evolution, the details of the actual process are still unknown. Therefore, in this study, we investigate orbital angular momentum ($J_{o}$), systemic mass ($M$) and orbital period ($P$) distributions of CAB and W UMa systems together.

\section{data}

Table 1 contains 119 CAB stars from \cite{K04} with basic data (orbital and physical). Columns are organized as order number, name, spectral type, stage of evolution, total mass ($M=M_{1}+M_{2}$), mass ratio ($q=M_{2}/M_{1}<1$), orbital period ($P$), primary radius ($R_{1}$), secondary radius ($R_{2}$), logarithm of OAM ($J_{o}$) and spin AM ($J_{spin}$) as a fraction of OAM. Rows are organized as first 53 stars are possible members of young moving groups (MG) and the rest 66 stars are field systems. Furthermore, rows are sorted in increasing orbital periods.

CAB stars in Table 1 had been classified into three stages of evolution G, SG and MS (column 4) according to their luminosity classes. One can see \cite{K04} for details. Simply, G group contains systems at least one component being a giant. SG group contains systems with at least one subgiant but no giants. MS marks the systems with both components are on the main sequence. Systems with unknown secondaries were classified according to primaries. 

A mean kinematical age of 3.86 Gyr was assigned to field CAB stars by \cite{K04} from their galactic space velocity dispersions. The field CAB group contains younger and older stars as a mixture. But, the other CAB group (MG) contains only the young ones. One may see \cite{K04} for the details how MG systems were selected from the common CAB. The ages of MG groups are known as open cluster ages by the turn off point from the main--sequence. Consequently, a pre--determined age of a moving group can be assigned to binaries which were found to be possible members according to their space velocity vectors. Therefore, unlike the field stars with various ages, the MG stars are homogeneous with a single age corresponding to each MG. Among the five MG groups considered by \cite{K04}, the Hyades super cluster is the oldest one with 0.6 Gyr age \citep[see Table 3 of][]{K04}. Therefore, MG systems are considered to be younger than 0.6 Gyr. Kinematical criteria of MG were defined by \cite{E58a, E58b, E89, E95} and summarized by \cite{M01a, M01b}.

Table 2 contains 102 W UMa stars from \cite{B05} with basic data (orbital and physical). Columns are organized the same as Table 1 with only one exception that column 4 contains the type rather than the evolutionary stage. There are two types of W UMa systems according to the light and the velocity curves \citep{B70}. W-type systems are those such as W UMa itself, in which the hotter component (the star eclipsed at the primary minimum) is smaller and less massive. Rows are organized as first 26 with MG designations and the rest 76 stars are field systems. As in Table 1, rows are sorted in increasing orbital periods.	

The spin AM ($J_{spin}$) of binaries is less well known than the orbital AM because the radii of gyration are uncertain. They are needed if one wants to study total AM. They are computed for this study just to compare them to OAM. An analytical approximation and coefficient from \cite{CG90} for main-sequence stars were used in computing assuming synchronous rotations whenever the radii of both components are available. However, since their approximation is for the mass range between 0.60 and 25 $M_{\odot}$, for some CAB and W UMa stars with component masses less than 0.60 $M_{\odot}$, $J_{spin}$ has not been computed. On the other hand, the computed $J_{spin}$ for evolved stars has to be considered as upper limits because the moment of inertia is known to be decreasing by evolution. Those rough values of the total spin AM are displayed as fractions of OAM (as $J_{spin}/J_{o}$) in Table 1 and Table 2. It has been found that total spin AM of all systems (CAB and W UMa) usually less than one tenth of $J_{o}$. There are only nine systems, which are all contact systems (AW UMa, $\epsilon$ Cra, FG Hya, OU Ser, TZ Boo, TV Mus, V410 Aur, V776 Cas, XY Boo) with total Spin AM higher than one tenth but less a quarter of $J_{o}$. In average, W UMa systems have nearly 10 times more spin AM than those of CAB stars. 

In this study, we consider orbital dynamics, e.g. OAM changes. If AM transfer occurs between the orbit and the spinning components due to tidal interactions, this study considers it as an instability in the orbit. Any AM transfered from the orbit to the component stars will be sensed by the orbit as an OAM loss and vice versa. For the rest of this study, $J_{o}$ strictly mean OAM, one must not confuse it with the total AM which includes both orbital and spin AM existing in a binary.

\section{Orbital dynamics}
The most basic definition of OAM ($J_{o}$) can be given as

\begin{eqnarray}
J_{o}={M_{1}M_{2}\overwithdelims()M_{1} + M_{2}}{a^2}\Omega=
{q \overwithdelims()(1 + q)^{2}}{M a^2}\Omega.
\end{eqnarray}
where $I={M_{1}M_{2}\overwithdelims()M_{1} + M_{2}}{a^2}={q \overwithdelims()(1 + q)^{2}}{M a^2}$
is moment of inertia and $\Omega = 2\pi/P$ is angular speed for an orbital motion, thus $J_{o}=I\Omega$.

$J_{o}$ and $M$ are two basic physical quantities which determine a unique period ($P$) and a unique size for the orbit as

\begin{eqnarray}
P={(1+q)^{6} \over q^3} {2\pi \over G^2}{ J_{o}^3\over M^5},\qquad\qquad a={(1+q)^{4} \over Gq^2} { J_{o}^2\over M^3},
\end{eqnarray} 
where the mass ratio ($q = M_{2}/M_{1} < 1$) can be considered as an auxiliary parameter used in the definition of $J_{o}$. The size $a=a_1+a_2$ represents the semi--major axis of a  relative orbit of one star around the other. Stability of an orbit ($dP=0$, $da=0$) requires $J_{o}$ and $M$ to be constant ($dJ_{o}=0$, $dM=0$) provided with no mass transfer ($dq=0$). If there is no mass transfer, which must be true for detached binaries, it is obvious that OAM loss will cause an orbit to reduce its period and size. On the contrary, mass loss has an affect of increasing the period and the size. Logarithmic derivatives of (2) give all possible relative changes as

\begin{eqnarray}
{dP \over P} = -3 {{1-q}\over{1+q}} {dq\over q} +3 {dJ_{o} \over J_{o}}-5{dM\over M},\qquad\qquad {da\over a} = -2 {{1-q}\over{1+q}} {dq\over q} +2 {dJ_{o} \over J_{o}}-3{dM\over M}.
\end{eqnarray} 
Because $M$ has higher power than $J_{o}$, the affect of mass loss would dominate. For example, in the case of same relative changes of OAM and mass (isotropic stellar winds, if $dq=0$), $dJ_{o}/J_{o}=dM/M$ according to (1), then

\begin{eqnarray}
{dP \over P} = -2{dJ_{o} \over J_{o}}=-2{dM\over M},\qquad\qquad {da\over a}=-{dJ_{o} \over J_{o}}=-{dM\over M}.
\end{eqnarray} 
which means mass loss and corresponding OAM loss will have a net effect on the orbit to increase both the period and the size. However, there could be additional causes to increase relative OAM loss, e.g. OAM loss of  gravity waves, or stellar encounters in the galactic space, or a third body in an eccentric orbit around the binary system, or existence of an amplification mechanism such as in some tidally locked binaries, where tidal interactions transfer OAM to spinning components and transferred AM is lost at the Alfven radius. After considering all possibilities, one has to compare the grant total relative OAM loss to the relative mass loss which could be expressed by a parameter $\delta$ defined as 
\begin{eqnarray}
{\delta} = {({dJ_{o} \over J_{o}}) / ({dM \over M})},
\end{eqnarray}
which can be called dynamical parameter because dynamical respond of the orbit depends on the value of $\delta$. It should be noted that $\delta$ is equal to the amplification parameter $A=(dJ_{o}/dM)/(J_{o}/M)$ as defined by \cite{D06}.  Inserting $\delta$ into (3),

\begin{eqnarray}
{dP \over P} = (3-{5\over \delta}){dJ_{o} \over J_{o}}=(3\delta -5){dM\over M},\qquad\qquad {da\over a} = (2-{3\over \delta}){dJ_{o} \over J_{o}}=(2\delta -3){dM\over M}.
\end{eqnarray}
Change in mass ratio could be negligible because of a comparable relative mass loss from the component stars. Even if $dq/q \neq 0$, the term $(1-q)/(1+q)$ could be very small especially for high mass ratio ($q\sim1$) systems. Thus, ignoring it in the first approximation is acceptable. 
 
Consequently, in eq. (6), $\delta>5/3$ is required to decrease the period. But, $\delta > 3/2$ is sufficient to shrink an orbit. If $3/2<\delta<5/3$, orbital size decreases despite period is increasing. The size and the period of an orbit both increase if $\delta < 3/2$.

Using the mean decreasing rates of OAM and mass from  \cite{D06}, the mean value for the dynamical parameter ($\bar{\delta}$) for detached CAB systems can be estimated as
\begin{eqnarray}
{\bar {\delta}} = {{dJ_{o} \over J_{o}} \over {dM \over M}} = {{dJ_{o} \over J_{o}dt} \over {dM \over Mdt}} = {-3.48 \times 10^{-10} \over -1.30 \times 10^{-10}}=2.68 
\end{eqnarray}
in the solar neighborhood.

There may be alternative AM loss mechanisms among the detached CAB. Direct loss of OAM is always possible in the galactic space due to stellar encounters, although it would be negligible for short period systems \citep{S95, GNM93}. Close encounters are more likely in multiple systems during periastron passages of distant third companions, which can lead to a rapid AM loss from the smaller orbits \citep{KEM98, EK01}. Therefore, we have preferred to use $\bar \delta = \bar A = 2.68$ of \cite{D06} as the mean dynamical parameter, which refers to all possible OAM loss mechanisms when describing the mean dynamical evolution on $\log J_{o} - \log P$, $\log M - \log P$, $\log J_{o} - \log M$ diagrams in the following sections.        
   
\section{discussions}
\subsection{The $\log J_{o}-\log P$ and $\log M-\log P$ diagrams}

The orbital angular momenta ($J_{o}$) and the periods ($P$) of CAB (Table 1) and W UMa (Table 2) systems are all plotted on a $\log J_{o}-\log P$ diagram in Fig. 1. CAB stars containing G, SG and MS designations and A \& W types of W UMa stars are indicated. With smaller orbital periods ($P<$1 day) W UMa stars are concentrated at the lower left while CAB on the right display a wider band elongated towards the upper right. The constant total mass ($M=M_{1}+M_{2}$) lines are computed using
\begin{eqnarray}
J_{o}={q\over (1+q)^2}\sqrt[3]{{G^2 \over 2\pi}M^5P},
\end{eqnarray} 
where $P$ is varied while a chosen $M$ is fixed. To represent a typical CAB, $q=0.88$ median value for the present sample were used.

The G systems with the biggest total masses prefer longer orbital periods and thus their OAMs are larger. The MS systems  with smallest masses and shortest orbital periods have comparable OAMs with W UMa systems. Having moderate masses and orbital periods, the SG systems hold moderate OAMs. Unlike CAB with S, SG and MS designations, A and W type W UMa stars do not have distinct locations since they appear totally mixed. 

A well defined smooth upper boundary of CAB stars appears as if tracing a path of dynamical evolution. OAM loss, mass loss and associated orbital period decrease would move a system form the upper right to the lower left parallel to the upper boundary. Finally some systems would enter in the region of contact binaries. With a similar idea and similar data set of less number of stars, \cite{D99} has computed $dJ_{o}/dP$ directly from the inclination of the upper boundary line and estimated $dP/dM$ and $dJ_{o}/dM$ from its cut positions by the constant total mass lines. Decreasing of the total masses on the upper boundary had been interpreted as the mass loss which activates OAM loss. However, a logical reason why a system moves parallel to the upper boundary is not clear. The $\log J_{o}- \log P$  diagram gives no clues on the time derivatives ($dJ_{o}/dt$, $dP/dt$ and $dM/dt$) either. 

It is easy to explain why lower right of CAB distribution on Fig. 1 is empty by selection effects. Being less bright , small--mass, long--period systems could be missed by observers or have not yet studied. Such an excuse, however, does not exist to explain absence of short--period but more massive CAB systems on the diagram (upper left of CAB distribution). Therefore, we are inclined to think such systems do not exist at all. Otherwise, they would have been noticed and observed as some of the CAB stars in our list.

In order to investigate why there exist such a well defined upper boundary, the $\log M - \log P$ and $\log J_{o}- \log P$ diagrams are compared in Fig. 2. Similar distribution characteristics are observable on both diagrams that the smooth upper boundary also exist on the $\log M - \log P$ diagram. First, the upper boundary of CAB sample on the $\log M - \log P$ diagram was eye estimated and digitized by computer. The estimated equation
\begin{eqnarray}
\log M= 0.155\log P+0.399 
\end{eqnarray}
seems to mark upper mass limits for various orbital periods.

Next, from the $P$ and $M$ values of this line, we have computed the corresponding line on the $\log J_{o} - \log P$ diagram using equation (8) with $q=0.88$. Equally well, perhaps better fitting of the re-produced line on the $\log J_{o} - \log P$ diagram clearly declares that the upper boundaries of these two diagrams are not independent and appears to be determined by mass upper limits for various orbital periods. Could it be that the mass upper limits imply initial Roche lobe structures with sizes primarily depend on the sizes of the preliminary orbits, where the mass within the lobes become limited according to the density of the star forming regions? or, is it only due to mass--period--activity relation? Why is the upper boundary not parallel to the constant total mass lines? (see Fig. 1). Special investigations other than this study seem to be needed. 

If period decrease occurs because of OAM loss but no mass loss and transfer, a dynamical evolution would follow a path parallel to the constant total mass lines in Fig. 1, which would carry CAB stars into the empty region. Either this is not happening or systems moving into the region seize chromospheric activity. Therefore, the CAB upper boundary on $\log J_{o} - \log P$ diagram may indicate a dynamical evolution with a minimum mass loss. Nevertheless, mean decreases determined from the kinematical ages of CAB by \cite{D06}, which are shown by the right sides of the triangles in both diagrams in Fig. 2, indicate that the direction of the mean dynamical evolution neither is parallel nor towards the empty region as it is shown by the arrows at the upper ends of the upper boundaries  in both diagrams in Fig. 2.

Both diagrams in Fig. 2 have been plotted intentionally to indicate MG and field systems, in order see aging effect. Unlike the sub--groups with G, SG and MS designations, which prefers certain locations, kinematically young (ages $<$ 0.6 Gyrs) MG systems appear randomly mixed among the kinematically old (3.86 Gyrs) field CAB. Random distribution of young and old systems in the CAB region confirms that an active binary can be born or may start its dynamical evolution anywhere on those diagrams. That is, a same location could be occupied equally likely by an old or a young system. Since there is no definite starting reference point (note: the direction and the speed may be known), it is not possible to trace the dynamical evolution of binary orbits on those plots in a similar manner as a single star nuclear evolution is traced on the H-R diagram.   

Unlike, MG and field CAB, which has no preferred location on the $\log J_{o} - \log P$ and $\log M - \log P$ diagrams, the W UMa stars designated with MG and field in Fig. 2 seem to show distinct locations which will be discussed in the following.  

\subsection{$\log M-\log P$ and $\log J_{o}-\log P$ distributions of W UMa systems}

W UMa systems do not display similar distribution characteristics as CAB systems. Downward curvature of the upper boundary at the left end of Fig. 1 (W UMa region) implies a decrease of OAM because of hiding considerable fraction of total AM as a spin AM since synchronous rotation rates increase towards the shorter periods. Displaying the W UMa region in a larger scale, Fig. 3 allows to compare mass and OAM distributions. Not only OAM distribution but also total mass distribution shows a similar trend. Therefore, the downward curvature of the upper boundary of W UMa region cannot be entirely due to transforming considerable OAM to spin AM. It is more likely to be related to mass content within the Roche lobes as the $\log M-\log P$ diagram suggests. At smaller orbital periods, orbits are smaller. Smaller Roche lobes contain less mass. One should also not forget OAM is more sensitive to orbital sizes than the total masses and periods as displayed in equation (1).

The upper boundary of CAB region was extrapolated into the W UMa region and displayed as solid and dashed lines in Fig. 3. One can easily notice that there are considerable number of contact systems over those lines especially at upper right. Those are the systems with excess mass not only with respect to the rest of the W UMa sample but also with respect to the mass-period distribution inferred from the CAB systems. With an excess mass for a given orbital period, it is easier to fill Roche lobes in the early stages of formation.             

There are 18 systems above the solid line in Fig. 3a and half of them are with MG designations. Stars with MG designations have relatively larger masses and longer orbital periods with respect to the rest of the W UMa sample. Another interesting feature of W UMa distribution on Fig. 3a is that there is a square shaped empty region on the lower right corner that longer period systems ($\log P > -0.35$) all have total masses bigger than 1.9 $M_{\odot}$ ($\log M>0.28$). 

Displaying an unusual distribution with respect to the rest of the W UMa sample, the 26 systems framed in a rectangle in Fig. 3 took our attention as young systems which were possibly born as contact binaries (with no pre--contact detached phases). We have restudied the galactic space velocities of those framed 26 systems, inspected their $U$-$V$ diagrams, and computed their kinematical ages. Having the dispersions of 26.10, 19.16, 19.18 kms$^{-1}$ at $U$, $V$, $W$ and 37.63 kms$^{-1}$ at the combined total space velocity vector $S=\sqrt{(U^2+V^2+W^2)}$, where $U$, $V$, $W$ are galactic space velocity components with respect to LSR (Local Standard of Rest), an average age of 2.47 Gyr were found for them. It is also interesting that 15 out of 26 have positive $V$ velocities bigger than Sun's $V$ velocity in the LSR, which constitutes an additional strong argument favoring them to be young. Only two systems, $\epsilon$ Cra and V2388 Oph, appear kinematically different. If these two systems were excluded, the average kinematical age reduces to 2.00 Gyr. 

Major contribution to the OAM loss at the contact stage comes from the gravitational wave radiation \citep{GB88}, which is inversely proportional to the orbital period. Having the largest orbital periods with respect to the rest of the W UMa sample, the systems in the frame are expected to be less effected by gravity waves thus; their lifetime on the contact stage must be longer. According to \cite{B05}, average lifetime of the contact stage is 1.61 Gyr, which is already slightly longer than the lifetime ($0.1 < t_{contact} < 1$ Gyr) according to \cite{GB88}. Thus, the 2.00 Gyr age found for the systems in the rectangle strongly favors them as contact binaries which were born contact. Existence of a flat lower boundary for the framed systems could be a hint for this. The systems below the flat boundary (systems with masses less than $1.9M_{\odot}$), apparently do not had sufficient masses to be formed as contact systems. The systems below must have been formed as detached systems and after their first Roche lobe overflow, their orbital periods decreased quickly due to mass transfer and they are now located among the other less massive contact systems. 

In fact, the region is not really empty but scarcely populated by detached systems (see Fig. 1 and Fig. 2). We believe those are the detached systems still evolving dynamically from the detached stage to the contact stage by loosing mass and orbital angular momentum. Before reaching a state of full contact, a mass transfer from more massive primary to less massive secondary must occur first. Such a mass transfer is not commonly observed because it is a fast stage at which the system mass and angular momentum appears to be conserved while the period of the orbit decreases quickly. If it is not possible to reach a full contact configuration in this stage, a detached binary may lose a great deal of its chance to be a contact binary and may become a classical Algol because mass loosing primary becomes less massive, then it is recognized as the mass loosing secondary. Mass transfer from the secondary to primary is a rather slow process which forces orbital period and size to be bigger. Only if there is sufficient OAM loss, the orbital period may turn to decreasing.

W type W UMa systems dominate over A types towards the shorter periods (see Fig. 3). Therefore, according to the mass transfer evolution of an orbit as described above, W systems seems to be the first formed at the first mass transfer stage. Then, during the slow mass transfer stage after the mass ratio reversal, their orbital periods start to increase slowly and systems appear moving towards the longer periods into the region of A type systems. Existence of many W systems with increasing orbital periods \cite{Q03} seems to confirm this scenario. A types being formed after W types have been suggested by \cite{W78}, \cite{M81}, and \cite{AH05}.  

Just recently, \cite{GN06}, who studied mass-period distribution of over 100 W UMa stars, have noticed A type systems usually have larger total masses. It is not very obvious but one can see Fig. 3b may also confirm this. According to \cite{YE05}, near contact binaries can transform to A type W UMa stars without a need of mass loss. \cite{GN06} propose A type systems may transform to W type systems with a simultaneous mass loss. The opposite were claimed unreasonable since it requires the total masses to increase \citep{GN06}. However, it is clear in Fig. 3b that there are some A type systems with total masses less than the total masses of some W type systems. For a massive W system, there still could be chance to transform to a less massive A type system. The two ways of forming A type systems cannot be ruled out only by comparing their mean masses.     

Do both A type and W type contact systems have a common origin? Is the one type the progenitor of the other? Such questions are still not answered. Formation of W UMa systems will be considered in the following section again when dynamical evolution from detached to contact state becomes clearer.            
 
\subsection{The $\log J_{o}–-\log M$ diagram}
Because OAM ($J_{o}$) and mass ($M$) are basic physical quantities determining orbital size and period ($a$ \& $P$), and because OAM loss and mass loss are physical parameters controlling the magnitude and direction of dynamical evolution, the $\log J_{o}-–\log M$ diagram is a natural choice to study dynamical evolution of binary orbits. Once, the diagram (Fig. 4) is produced, a sharp separation between the detached and contact systems stroke to our attention. Goodness of the separation is out striking that despite crowding along the border, there are only two systems (OO Aql, $\delta$ Cap) on the wrong side, which could be due to a wrong identification of the state of being contact or just because of observational errors. Similar separation does not occur on the diagrams discussed before.

Marking several positions on the borderline between CAB and W UMa stars, the following quadratic equation was produced.
\begin{eqnarray}
\log J_{lim}=0.522(\log M)^{2}+1.664(\log M)+51.315,
\end{eqnarray}  
where $M$ is in solar units and $J_{lim}$ is in cgs. Physical significance of this line is that it marks the maximum OAM for a contact system to survive. It is like in single stars, spin AM has to be less than a certain value otherwise gravity cannot hold stellar mass together. If OAM of a contact system is more than $J_{lim}$, the contact configuration brakes. 

Because W UMa systems all have circular orbits (CAB systems of small orbital periods too) and because OAM (eq. 1) is more sensitive to component separations than periods and masses, the line determined by (10) can be called ``the contact border''. This is also because if it were possible to increase ($J_{o}$) of a contact system, there will be a limiting value associated with a limiting separation of components. Any $J_{o}$ bigger than $J_{lim}$, the distance between the components would be bigger than the limiting distance permitting a full contact. 

CAB systems (RT Lac, AR Mon, $\epsilon$ UMi, RV Lib, BH CVn) eliminated from the list of \cite{D06}, since they are filling or about to fill one of the Roche lobes that mass transfer possibly occurring in them, are marked on Fig. 5. Since those systems are not close to the contact border and scattered randomly all over in the detached region, semi--contact configuration, like the other detached states, does not guarantee being close to the contact border.    

If total AM rather than OAM were plotted on Fig. 4, the smoothness of the border would have been spoiled because as discussed by \cite{R95} for low $q$ contact binaries, the total spin AM of the primaries may be comparable with OAM, so points of low $q$ systems would be lifted by up to $1/3$ of its value. It would have been meaningless to search such a border on a diagram using total AM. Moreover, adding total spin AM to OAM would have introduced additional uncertainties since spin AM is less certain than OAM because radii of gyration are uncertain. So it is not a coincidence for us to notice the contact border on the $\log J_{o}-–\log P$ diagram.  

Other diagrams ($\log J_{o}-–\log P$, $\log M-–\log P$) do not have such a well defined region and borderline for the contact binaries because according to equation (2) additional parameter ($q$) is needed to transform $J_{lim}(M)$ function into $P_{lim}(M)$ function as
\begin{eqnarray}
P_{lim}=0.0046{(1+q)^{6} \over q^3}M^{1.566\log M-0.008},
\end{eqnarray} 
where $P_{lim}$ is in units of days if $M$ is in solar mass. It is also possible to express limiting separation of components as
\begin{eqnarray}
a_{lim}=0.1167{(1+q)^{4} \over q^2} M^{1.044\log M+0.328}, 
\end{eqnarray} 
where $a_{lim}$ is in the units of solar radius if $M$ is in solar mass. Because limiting periods (or separations) strongly depends on ($q$), a unique transformation of the contact border into the other diagrams is not possible. As displayed in Fig. 1 and Fig. 2, there are detached CAB systems with periods less than one day occupying the same region with W UMa stars on both  $\log J_{o}-–\log P$ and $\log M -–\log P$ diagrams. With its own $q$, each system has its own $P_{lim}(M)$. Therefore, mixing CAB and W UMa stars is unavoidable on those diagrams.

Being able to confine detached and contact systems into separate regions, $\log J_{o}-–\log M$ diagram is best for studying the dynamical evolution of binary orbits. The diagram clearly displays that there is a well defined possibility for a detached system to go into the region of contacts by losing OAM. This possibility, however, depends on systems position on the diagram as well as speed and direction of its dynamical evolution, which could be very different from one system to another. This diagram too, like other diagrams, gives no clue on individual dynamical evolutions since there is no information about initial positions. Random mixing of young (MG) and old (field) systems also occurs in this diagram. That is, a same position could be belong to both a young and an old system. 

Nevertheless, the contact border (eq. 10) could be helpful to determine contact binary candidates. The amount of mean losses ($\Delta J$ and $\Delta M$) corresponding to 2, 4 and 6 Gyr are subtracted from the $J_{o}$ and $M$ values of the border. That is, the contact border is shifted accordingly. Dotted lines in Fig. 5 represent shifted borders that the systems between a dotted line and the contact border are the ones, which have a definite chance to reach at the contact border within the time intervals indicated if their nuclear evolutions permit them to live as much. After counting, it becomes clear that 11, 23 and 39 per cent of the current sample of CAB could pass over the contact border within the next 2, 4 and 6 Gyr according to the mean dynamical evolution with $\bar \delta=2.68$.    

The dynamical evolution of contact binaries is more uncertain than detached CAB stars since a mean $\delta$ is not available for them. If there is a smooth transition from the detached to the contact region, W systems are first to form according to W and A type distributions on the  $\log J_{o}-–\log M$  diagram (see Fig. 4) because W systems are usually closer to the border than A type systems. However, forming first W types and then A types is not reasonable according to \cite{GN06} since this  would need a mass gain. If A types are first to form, a smooth transition into the contact region also becomes unreasonable. This is because, forming W types later than A types appears also problematic since this require OAM gain (see Fig. 4). We encourage theoretical studies to investigate possibilities of jumping from the detached region into the contact region to form A type W UMa stars, which may occur during the first mass transfer stage, and then to produce W type W UMa stars. A to W or W to A transitions, if occuring, must be done within a lifetime of a contact binary which is known to be very short ($0.1 < t_{contact} < 1$ Gyr according to \cite{GB88}, 1.61 Gyr according to \cite{B05}).      

\subsection{Period and size evolution of orbits}
It is possible to draw constant period lines on the $\log J_{o}-–\log M$ diagram using (8). On the other hand, 
\begin{eqnarray}
J_{o}={q\over (1+q)^2}\sqrt{GM^3a}
\end{eqnarray} 
can be used to compute constant orbital size lines similarly. From the statistics of present CAB and W UMa samples, the median values of $q=0.88$ and $q=0.39$ are found and used to represent CAB and W UMa systems in computing constant period and size lines which are shown in Fig. 6. The constant orbital period and orbital size lines run almost parallel to the contact border and both $P$ \& $a$ values decrease towards it. Further decrease into the region of contacts is also clear.    

Constant period and constant size lines are sensitive to small $q$ values. One can feel the sensitivity by comparing the lines of $P=1$ day (or $a=3R_{\odot}$) computed by $q=0.88$ and $q=0.39$, and  $P=0.2$ days (or $a=1.2R_{\odot}$) computed with $q=0.39$ and $q=0.1$. Decreasing periods (or sizes) towards the lower right is deceptive as if evolution to contact stage is occurring from upper left to lower right which is impossible since such an evolution requires a mass gain. Mass loss, however, changes the direction from vertically down (OAM loss only) towards to the lower left (if OAM and mass both are lost). Because mass loss dominates over OAM loss (eq. 3), there are lower limits; one for orbital periods and one for the orbital sizes. Both limits are indicated by the dotted lines in the triangles and the corresponding numerical values ($\Delta \log J_{o}/\Delta \log M$) in Fig. 6. Any dynamical evolution with a $\delta$ smaller than those limits indicates an increase rather than a decrease on both $P$ \& $a$. 

\section{Conclusions}
Total mass $M$ and orbital angular momentum OAM ($J_{o}$) are two basic physical quantities which determine sizes and periods of binary orbits. OAM has several different analytical formulae established with different combinations of mass ratio $q$, semi--major axis $a$, orbital period $P$ and systemic mass $M$. Because $M$ \& $J_{o}$ are well known basic quantities and  mass loss ($dM$) \& OAM loss ($dJ_{o}$) are again free parameters, from which the speed and direction of dynamical evolution can be inferred, the $\log J_{o}-\log M$ diagram is the best for studying the dynamical evolution of binary orbits. 

There are limits in orbital dynamics. $d\log J_{o}/d\log M>3/2$ is for decreasing orbital sizes. $d\log J_{o}/d\log M>5/3$ is for decreasing orbital periods. if $d\log J_{o}/d\log M$ takes a value between 3/2 and 5/3, orbital sizes shrink despite corresponding orbital periods increase. According to mean dynamical evolution with $\bar \delta=d\log J_{o}/d\log M=2.68$, as provided by \cite{D06}, the orbital sizes and periods of detached CAB systems are decreasing. 

A contact border, which separates W UMa stars from CAB, is discovered on the $\log J_{o}-\log M$ diagram. So, One can predict OAM of a system on the contact border, called $J_{lim}$, which is determined only by $M$. Any OAM less than $J_{lim}$ implies a full contact for a binary. Similar to $J_{lim}$, the limiting period $P_{lim}$, and limiting orbital size $a_{lim}$ are computable using empirical formulae (11) and (12). Unlike $J_{lim}$, which is independent of $q$, $P_{lim}$ and $a_{lim}$ have a strong dependence on $q$. 

Rather than computing the time left for each detached system to reach at the contact border, limiting borders corresponding 2, 4, and 6 Gyrs were computed by subtracting mean decreases ($\Delta J$ and $\Delta M$) of mean dynamical evolution from $J$ and $M$ values of the contact border. The CAB systems located between the contact border and those borders are the ones which are expected to be contact binaries within the time scales indicated. 

A detached system heading towards the contact region must go trough mass transfer stages before becoming a contact binary. At any mass transfer stage, even if it is conservative, OAM $J_{o}$ is not conserved. For shrinking orbits, $J_{o}$ must decrease according to eq. (1). For conservative cases, the OAM difference between the two different size orbits of same system are to be compensated by spin AM. That is, $J_{o}$ still decreases while the total spin AM is increasing. If a detach system is sufficiently close to the contact border, during the first fast mass transfer stage, it has a changes to jump over into the region of contacts and to be one of them.      

The traditional view of forming W UMa stars assumes that they only come from detached binaries of comparable periods. But, statistics available is convincing enough that there are very few such progenitors even to account for the low space density (0.2 per cent) of contact binaries in the solar neighbourhood \citep{R02, R06}. Righteously \cite{P06} says ``at this time the contact systems seem to appear OUT OF NOWHERE'' in order to emphasize the insufficiency of the traditional view. \cite{P06}, therefore, consider mechanical tree--body orbital evolution, which is a similar formation model of close binaries in globular clusters \citep{P03}, where the inner binary has relatively longer orbital period and has a better chance to evolve into a contact system after all \cite{PR06} found that up to 50 per cent of W UMa binaries have companions.

The mean dynamical parameter $\bar \delta$ used in this study naturally contains affects of all mechanisms of OAM loss since its value is derived from stellar kinematics. Perhaps, it is not possible now to make a firm conclusion which mechanism is more efficient before a complete statistics and their analysis are available. Nevertheless, \cite{B05} provided observational evidence that near contact detached binaries of few days orbital period are not the only source to form W UMa stars. They may also be born directly as contact systems. Moreover, it has also been found in this study that  considerable fraction of current detached CAB sample (39 per cent within 6 Gyrs) to have a potential to jump over the contact border on the $\log J_{o}-\log M$ diagram and become contact systems within a time scale of main-sequence life time. So, we encourage statistical studies to include these new findings after all contact binaries are rare with the local space density of just 0.2 per cent of the main sequence stars \citep{R02, R06}. 

\section{Acknowledgments}
We acknowledge the partial support by Turkish Scientific and Technical Research Council (T\"UB\. ITAK), \c{C}anakkale Onsekiz Mart University and Istanbul University. This study has been produced within the scopes of projects 104T508 of T\"UB\. ITAK and BAP 2005/108 of \c{C}anakkale Onsekiz Mart University. We acknowledge anonymous referee who took our attentions to critical points.

\begin{figure*}
\center
\resizebox{17cm}{9.58cm}{\includegraphics*{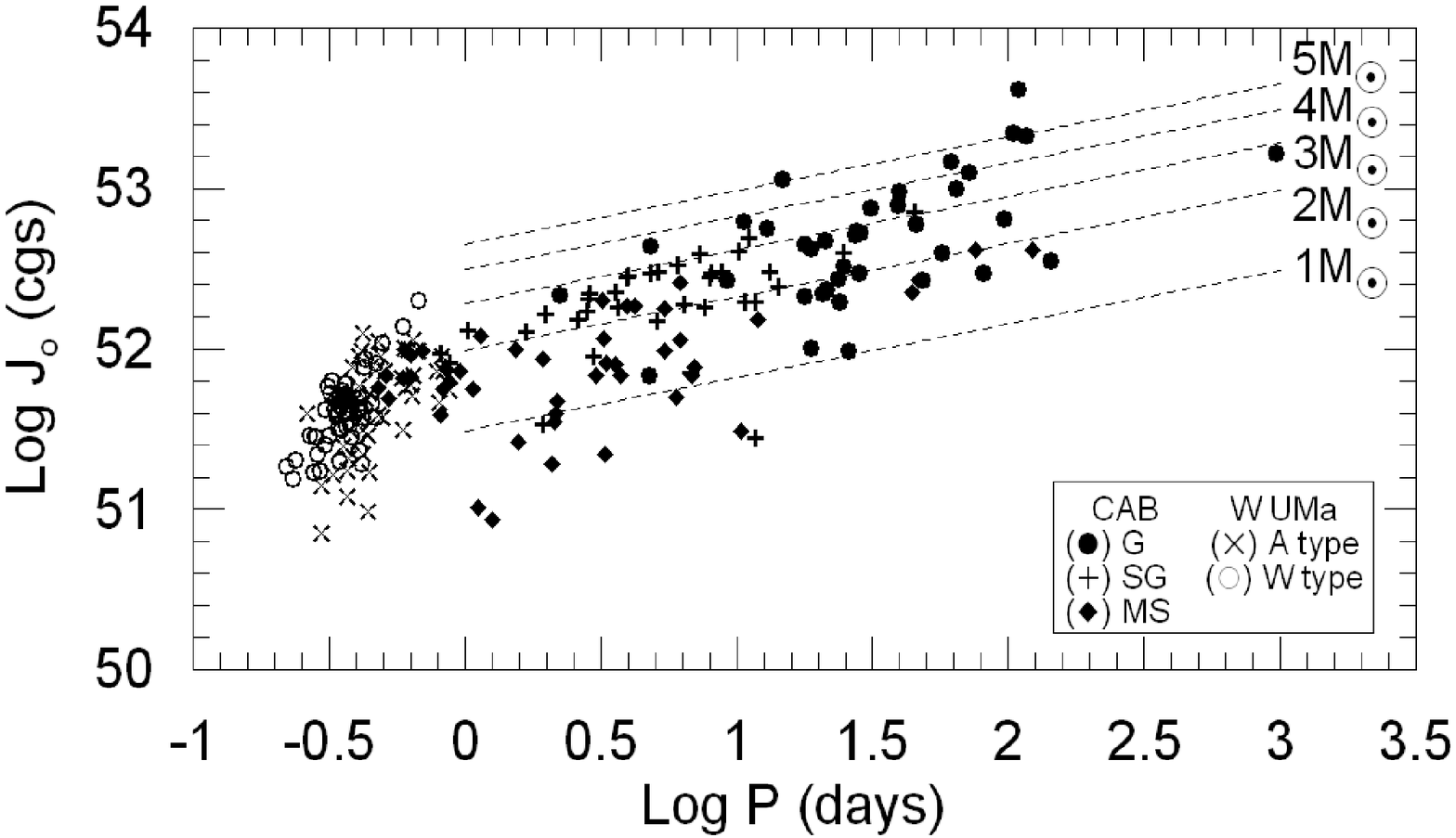}} 
\caption{Orbital AM and period distributions of CAB and W UMa stars are compared. Constant total mass lines (dashed) were computed using q=0.88, which is the median value for the present sample of CAB.}
\end{figure*}

\begin{figure*}
\center
\resizebox{16cm}{15.48cm}{\includegraphics*{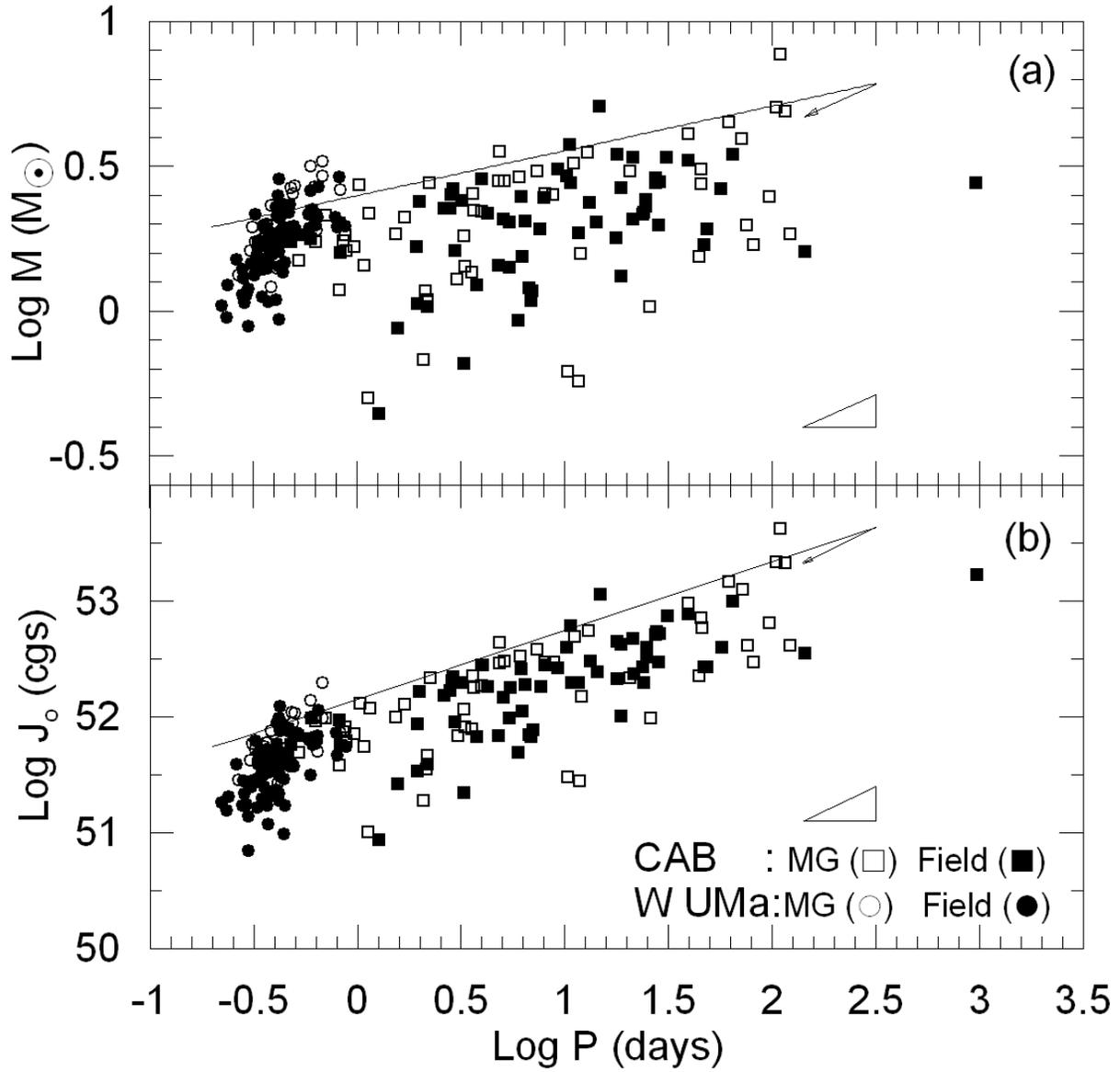}} 
\caption{Total mass--period (a) and OAM--Period (b) distributions are compared. CAB upper boundaries in both diagrams are not independent since it is possible to compute one from the other analytically. Young (MG) and old (field) CAB systems mix randomly. Mean dynamical evolution (arrow) and mean decreases (the right sides of triangles) are for 2 Gyrs.}
\end{figure*}

\begin{figure*}
\center
\resizebox{16cm}{22.34cm}{\includegraphics*{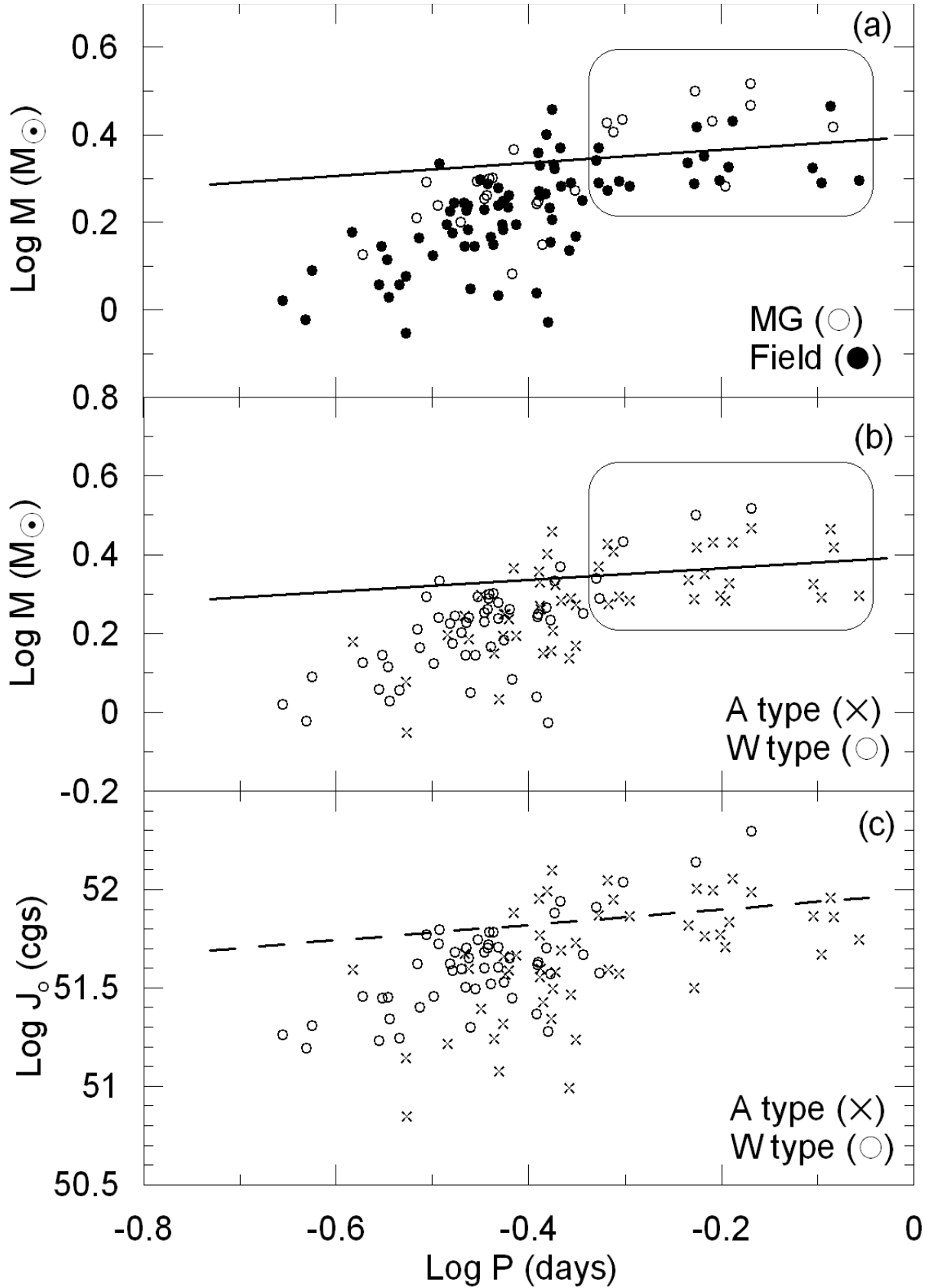}} 
\caption{W UMa regions of Fig. 1 and Fig. 2 are zoomed. Lines are CAB upper boundaries extended into W UMa region. Dashed is computed from solid using $q=0.39$ (median value of W UMa sample). Young kinematical age (2 Gyr) and the empty region below imply systems in the rectangle may be formed directly as contact binaries (born in contact).}
\end{figure*}

\begin{figure*}
\center
\resizebox{12cm}{23.15cm}{\includegraphics*{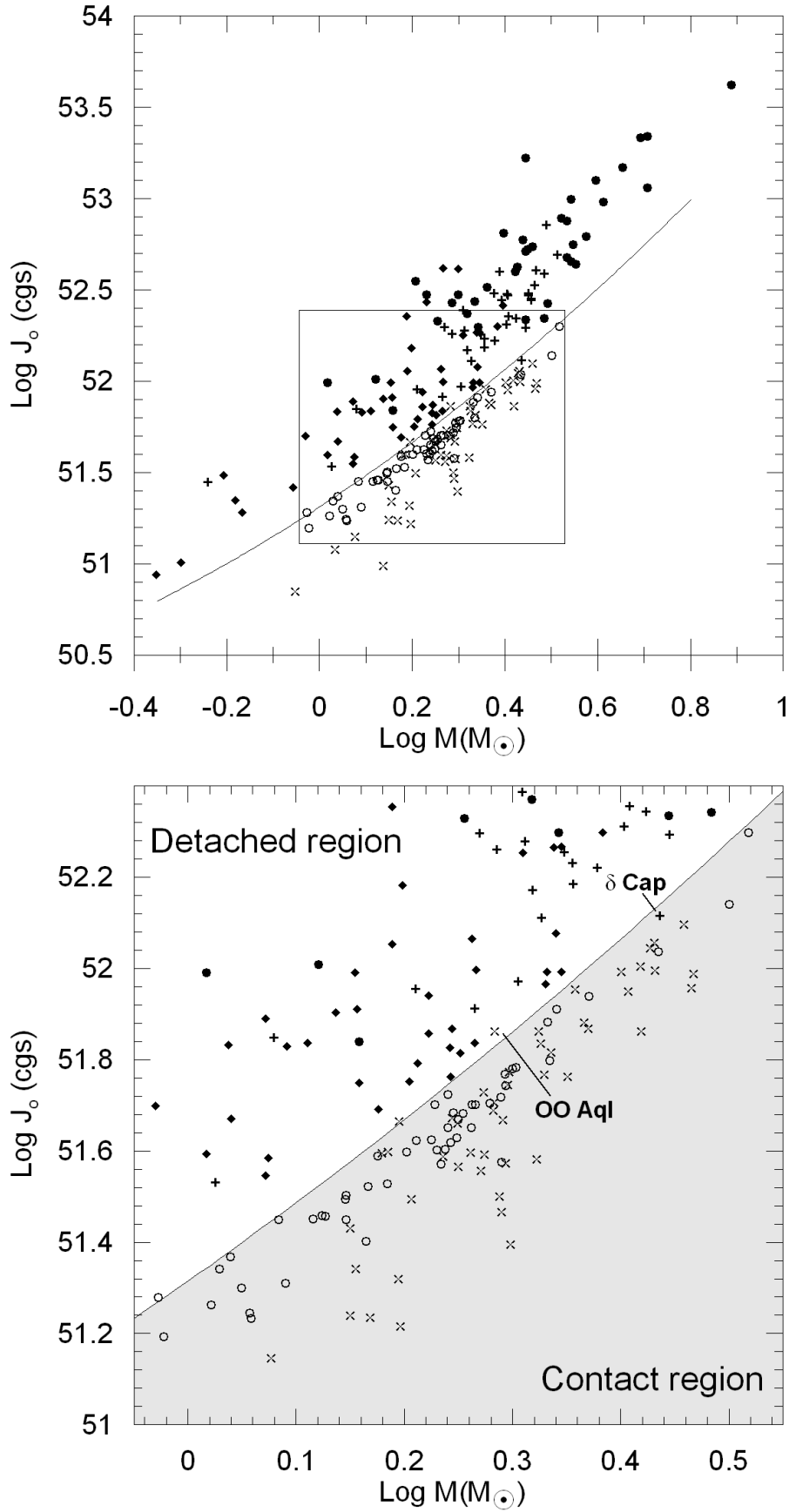}} 
\caption{A well defined borderline sharply separates detached and contact systems. Crowded region above (framed) is zoomed below. Symbols are like Fig. 1.}
\end{figure*}

\begin{figure*}
\center
\resizebox{12.52cm}{12cm}{\includegraphics*{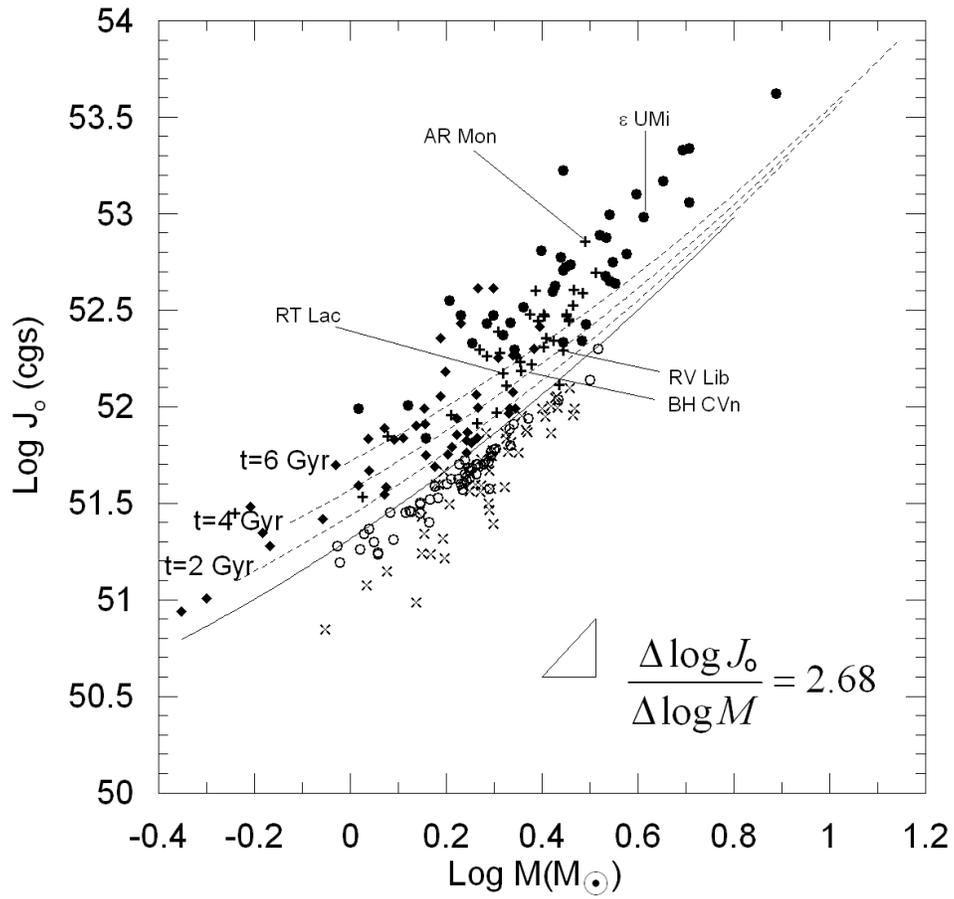}} 
\caption{Borders of equal times (dashed) to reach at the contact border (solid). Mean dynamical evolution (hypotenuse) and mean decreases (right sides) for 2 Gyrs.}
\end{figure*}

\begin{figure*}
\center
\resizebox{12cm}{21.69cm}{\includegraphics*{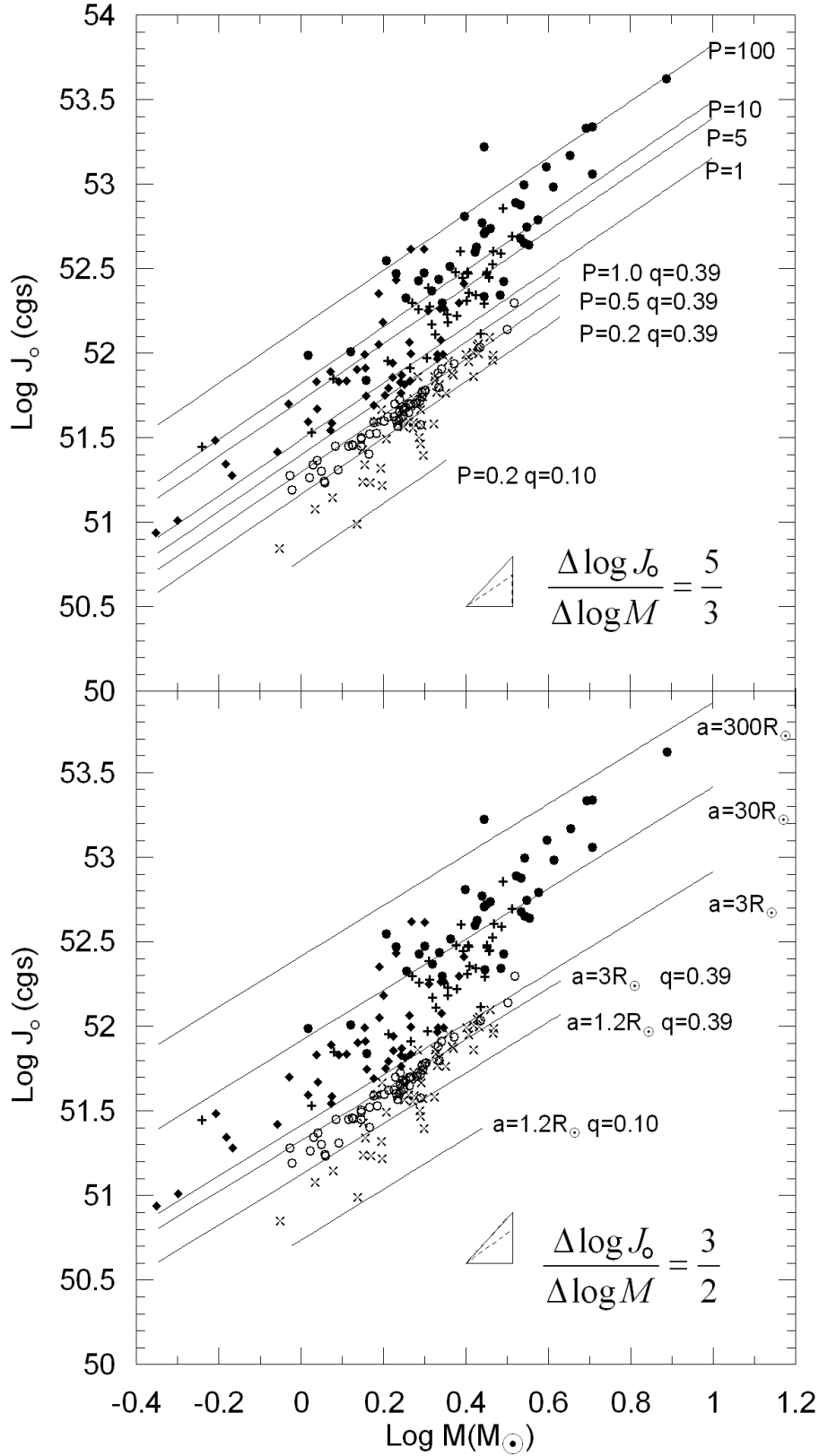}}
\caption{Constant period and size lines indicating Period and size evolutions of orbits. Mean (hypotenuse) evolution and corresponding decreases (right sides of triangles). Evolution with $\delta=5/3$ (dashed) keeps $P$ constant but evolution with $\delta=3/2$ (dashed) keeps $a$ constant.}
\end{figure*}

\begin{table*}
\caption{Physical parameters of the CABs. First 53 are possible members of young MG ($<$ 0.6 Gyr) and the rest are older (3.86 Gyr) field systems.}
\center
{\scriptsize
\center
\begin{tabular}{cllcccccccc}
\hline
        ID &       Name &   Spectrel       & Stage of &   $M_{tot}$ &    $q$     &    $P$    &  $R_{1}$ & $R_{2}$ & $\log J_{o}$ & $J_{spin}/J_{o}$\\
           &            &     Type         & Evolution&($M_{\odot}$)&            &    (days) &  ($R_{\odot}$) &($R_{\odot}$)& (cgs)& \\ 
\hline

         1 &   V471 Tau &     K2V+WD         &   MS &       1.50 &      0.974 &     0.521  &     0.83 & 0.01 & 51.690 & 0.017 \\
         2 &     RT And &    F8V+K0V         &   MS &       2.14 &      0.739 &     0.630  &     0.92 & 1.26 & 51.965 & 0.019 \\
         3 &     CG Cyg &      G9+K2         &   MS &       1.75 &      0.865 &     0.631  &     0.82 & 0.90 & 51.826 & 0.018 \\
         4 &     ER Vul &    G0V+G5V         &   MS &       2.15 &      0.957 &     0.698  &     1.08 & 1.11 & 51.992 & 0.014 \\
         5 &     YY Gem &  dM1e+dM1e         &   MS &       1.19 &      0.924 &     0.815  &     0.60 & 0.60 & 51.585 & 0.012 \\
         6 &     UV Psc &    G5V+K2V         &   MS &       1.75 &      0.765 &     0.861  &     0.83 & 1.11 & 51.869 & 0.013 \\
         7 &  V1430 Aql & G5V+K0III-IV       &   SG &       1.84 &      0.957 &     0.873  &     1.11 & 0.86 & 51.913 & 0.012 \\
         8 &   V772 Her & (G0V+?)+K7V        &   MS &       1.63 &      0.567 &     0.879  &     0.58 & 0.90 & 51.792 & 0.012 \\
         9 &     IL Com &    F8V+F8V         &   MS &       1.67 &      0.964 &     0.962  &     1.10 & 1.10 & 51.857 & 0.011 \\
        10 &$\delta$ Cap& F1IV-III/K1V       &   SG &       2.73 &      0.365 &     1.023  &          &      & 52.116 &       \\
        11 &     DH Leo & (K2V+K5V)+K5V      &   MS &       1.44 &      0.675 &     1.072  &     0.67 & 0.97 & 51.748 & 0.010 \\
        12 &    Gl 841A &     dM3-5e         &   MS &       0.50 &      0.917 &     1.122  &     0.34 & 0.36 & 51.008 & 0.008 \\
        13 &     TZ CrB &    F6V+G0V         &   MS &       2.19 &      0.975 &     1.140  &     1.10 & 1.14 & 52.077 & 0.008 \\
        14 & BD+23 2297 &    K1V+K1V         &   MS &       1.85 &      0.993 &     1.528  &     0.78 & 0.78 & 51.997 & 0.005 \\
        15 &   V824 Ara & G7IV/V+K0IV/V      &   SG &       2.12 &      0.909 &     1.683  &     1.42 & 1.55 & 52.110 & 0.006 \\
        16 &     13 Cet &    F8V+G4V         &   MS &       0.68 &      0.545 &     2.080  &          &      & 51.279 &       \\
        17 &   V478 Lyr &   G8V+dK-M         &   MS &       1.18 &      0.269 &     2.128  &     0.30 & 0.98 & 51.545 & 0.037 \\
        18 &     FF And &  dM1e+dM1e         &   MS &       1.10 &      0.970 &     2.173  &          &      & 51.670 &       \\
        19 &   V819 Her & (F2V+F8V)+G8IV-III &   G  &       2.78 &      0.704 &     2.228  &     1.29 & 1.87 & 52.335 & 0.004 \\
        20 &     KZ And &   dK2+dK2V         &   MS &       1.29 &      0.949 &     3.034  &          &      & 51.836 &       \\
        21 &BD +39 4529 &    F8V+K5V 	     &   MS &       1.83 &      0.564 &     3.243  &     0.64 & 1.10 & 52.064 & 0.002 \\
        22 &   V835 Her &    G8V+K7V         &   MS &       1.43 &      0.700 &     3.304  &     0.60 & 0.90 & 51.911 & 0.002 \\
        23 &     HZ Com &     G9+K4V         &   MS &       1.37 &      0.957 &     3.556  &     1.10 & 0.85 & 51.903 & 0.002 \\
        24 &     GK Hya &    F8+G8IV         &   SG &       2.56 &      0.910 &     3.589  &     3.39 & 1.51 & 52.356 & 0.004 \\
        25 &     UX Com &  K1(IV)+G2         &   SG &       2.23 &      0.855 &     3.639  &     2.50 & 1.00 & 52.255 & 0.003 \\
        26 &     BU 163 & (F9V/G0V)+?        &   MS &       2.22 &      0.947 &     3.963  &          &      & 52.266 &       \\
        27 &     RS CVn &  F6IV+G8IV         &   SG &       2.82 &      0.958 &     4.797  &     4.00 & 1.99 & 52.468 & 0.004 \\
        28 &     SS Cam & F5V-IV+K0IV-III    &   G  &       3.58 &      0.954 &     4.820  &     6.40 & 2.20 & 52.641 & 0.005 \\
        29 &     RT CrB &       G2IV         &   SG &       2.83 &      0.991 &     5.117  &     3.00 & 2.60 & 52.480 & 0.003 \\
        30 &     VV Mon &   G5V+G8IV         &   SG &       2.91 &      0.942 &     6.053  &     6.20 & 1.80 & 52.525 & 0.004 \\
        31 &     RW UMa &  F8IV+K0IV         &   SG &       3.06 &      0.951 &     7.328  &     4.24 & 2.31 & 52.588 & 0.002 \\
        32 &     LX Per & G0V-IV+K0IV        &   SG &       2.55 &      0.931 &     8.035  &     3.05 & 1.64 & 52.470 & 0.001 \\
        33 &     AW Her &  G2IV+K2IV         &   SG &       2.54 &      0.906 &     8.810  &     3.20 & 2.40 & 52.479 & 0.001 \\
        34 &  V1285 Aql &   dM2e+dMe         &   MS &       0.62 &      0.938 &    10.328  &     0.44 & 0.44 & 51.483 & 0.002 \\
        35 &     AE Lyn & F9IV-V+G5IV        &   SG &       3.25 &      0.979 &    11.066  &     2.64 & 3.14 & 52.693 & 0.001 \\
        36 &   V829 Cen &   G5V+K1IV         &   SG &       0.57 &      0.979 &    11.722  &          &      & 51.446 &       \\
        37 &   V808 Tau &    K3V+K3V         &   MS &       1.58 &      0.950 &    11.940  &     0.80 & 0.80 & 52.181 & 0.000 \\
        38 &     IL Hya & K1/2III/IV+G5V/IV  &   G  &       3.53 &      0.604 &    12.912  &          &      & 52.747 &       \\
        39 &  V1379 Aql &  K0III+sdB         &   G  &       3.05 &      0.129 &    20.654  &          &      & 52.342 &       \\
        40 & ADS 11060C &        K7V         &   G  &       1.04 &      0.951 &    25.763  &          &      & 51.990 &       \\
        41 &$\epsilon$ UMi& G5III+A8-F0V     &   G  &       4.10 &      0.464 &    39.446  &    12.00 & 1.70 & 52.982 & 0.001 \\
        42 & BD +64 487 &  K2-3V+K5V         &   MS &       1.54 &      0.921 &    44.361  &          &      & 52.354 &       \\
        43 &     KX Peg & F5-8V+G8IV         &   SG &       3.09 &      0.818 &    45.290  &          &      & 52.856 &       \\
        44 &BD +44 2760 &   G7III/GV         &   G  &       2.75 &      0.833 &    45.604  &          &      & 52.773 &       \\
        45 &     GT Mus & K2-4III+(A0)       &   G  &       4.50 &      0.800 &    61.376  &          &      & 53.171 &       \\
        46 &     DQ Leo & G5IV-III+A6V       &   G  &       3.95 &      0.879 &    71.614  &     5.90 & 1.70 & 53.101 & 0.000 \\
        47 & BD +17 703 &    G2V+G8V         &   MS &       1.99 &      0.881 &    75.683  &     1.00 & 1.00 & 52.614 & 0.000 \\
        48 &     BM Cam &      K0III         &   G  &       1.70 &      0.545 &    80.910  &          &      & 52.472 &       \\
        49 &      5 Cet &      K2III         &   G  &       2.50 &      0.786 &    96.383  &          &      & 52.810 &       \\
        50 &$\alpha$ Aur& G1III+K0III        &   G  &       5.09 &      0.951 &   103.992  &    12.80 & 8.70 & 53.341 & 0.000 \\
        51 &  V1817 Cyg & K2III-II+A0V       &   G  &       7.73 &      0.600 &   108.893  &          &      & 53.623 &       \\
        52 & $\eta$ And &   G8III/IV         &   G  &       4.93 &      0.904 &   115.611  &    11.00 &11.00 & 53.333 & 0.000 \\
        53 &  SAO 23511 &  F9.5V+G0V         &   MS &       1.85 &      0.682 &   122.180  &     0.75 & 1.00 & 52.616 & 0.000 \\
\\
        54 &     XY UMa &    G0V+K5V         &   MS &      1.748 &      0.606 &      0.479 &     1.16 & 0.63 & 51.762 & 0.022 \\
        55 &     BI Cet & G6V/IV+G6V/IV      &   MS &      1.840 &      0.916 &      0.515 &     0.90 & 0.90 & 51.836 & 0.024 \\
        56 &     SV Cam &    F5V+K0V         &   MS &      1.786 &      0.644 &      0.593 &     0.76 & 1.18 & 51.815 & 0.022 \\
        57 &     UV Leo &    GOV+G2V         &   MS &      2.214 &      0.970 &      0.600 &     1.19 & 1.08 & 51.992 & 0.019 \\
        58 &     BH Vir & F8V-IV+G2V         &   SG &      2.021 &      0.981 &      0.817 &     1.11 & 1.25 & 51.971 & 0.014 \\
        59 &     WY Cnc &  G0-8V+K2?         &   MS &      1.601 &      0.495 &      0.830 &     0.58 & 0.93 & 51.752 & 0.014 \\
        60 &     CM Dra &    M4V+M4V         &   MS &      0.444 &      0.926 &      1.268 &     0.23 & 0.25 & 50.938 & 0.005 \\
        61 &     CC Eri & K7.5V+M3.5V        &   MS &      0.876 &      0.537 &      1.563 &     0.41 & 0.64 & 51.419 & 0.011 \\
        62 &   V837 Tau &    G2V+K5V         &   MS &      1.670 &      0.673 &      1.932 &     0.74 & 1.00 & 51.941 & 0.004 \\
        63 &     EI Eri &       G5IV         &   SG &      1.060 &      0.379 &      1.945 &          &      & 51.531 &       \\
        64 &     AR Lac &  G2IV+K0IV         &   SG &      2.390 &      0.897 &      1.982 &     2.72 & 1.52 & 52.220 & 0.008 \\
        65 &     BK Psc &    K5V+M4V         &   MS &      1.040 &      0.552 &      2.168 &     0.45 & 0.72 & 51.594 & 0.010 \\
\end{tabular}  
}
\end{table*}

\begin{table*}
\contcaption{}
\center
{\scriptsize
\begin{tabular}{cllcccccccc}
\hline
        ID &       Name &   Spectrel       & Stage of &   $M_{tot}$ &    $q$     &    $P$    &  $R_{1}$ & $R_{2}$ & $\log J_{o}$ & $J_{spin}/J_{o}$\\
           &            &     Type         & Evolution&($M_{\odot}$)&            &    (days) &  ($R_{\odot}$) &($R_{\odot}$)& (cgs)& \\ 
\hline
        66 &     BH CVn &  F2IV+K2IV         &  SG &      2.271 &      0.544 &      2.612 &     3.27 & 3.10 & 52.184 & 0.008 \\
        67 &     CF Tuc &   G0V+K4IV         &  SG &      2.266 &      0.880 &      2.799 &     4.60 & 1.50 & 52.231 & 0.007 \\
        68 &   V711 Tau &  K1IV+G5IV         &  SG &      2.531 &      0.821 &      2.838 &     3.80 & 1.76 & 52.310 & 0.008 \\
        69 &     PW Her & K0IV+F8-G2         &  SG &      2.652 &      0.768 &      2.884 &     3.80 & 1.40 & 52.343 & 0.007 \\
        70 &     AD Cap & G5-8IV-V+G5        &  SG &      1.624 &      0.526 &      2.958 &          &      & 51.955 &       \\
        71 &     TY Pyx &  G5V+G5-8V         &  MS &      2.416 &      0.987 &      3.199 &     1.86 & 1.58 & 52.298 & 0.003 \\
        72 &  V1396 Cyg &   M2V+M4Ve         &  MS &      0.658 &      0.696 &      3.273 &          &      & 51.346 &       \\
        73 &     HZ Aqr &  K3Ve+K7Ve         &  MS &      1.234 &      0.804 &      3.758 &     0.45 & 0.55 & 51.830 & 0.002 \\
        74 &     SZ Psc &  K1IV+F8IV         &  SG &      2.861 &      0.766 &      3.963 &     5.10 & 1.50 & 52.444 & 0.006 \\
        75 &      Z Her &    K0IV+F5         &  SG &      2.864 &      0.843 &      3.990 &     2.73 & 1.85 & 52.450 & 0.003 \\
        76 & SAO 240653 &    G0V+G0V         &  MS &      2.180 &      0.998 &      4.236 &          &      & 52.264 &       \\
        77 &     UZ Lib &  K0III+A8?         &   G &      1.440 &      0.309 &      4.764 &    21.00 & 1.00 & 51.839 & 0.019 \\
        78 &     RT Lac &   G5V+G9IV         &  SG &      2.082 &      0.405 &      5.070 &     4.81 & 4.41 & 52.171 & 0.007 \\
        79 &     AS Dra &    G4V+G9V         &  MS &      1.426 &      0.889 &      5.408 &          &      & 51.991 &       \\
        80 &  V1423 Aql &    G5V+G5V         &  MS &      2.040 &      1.000 &      5.433 &          &      & 52.252 &       \\
        81 &     BY Dra & K6Ve + K7V         &  MS &      0.934 &      0.891 &      5.970 &          &      & 51.699 &       \\
        82 &     RS UMi &   G0V+G-KV         &  MS &      2.484 &      0.981 &      6.166 &          &      & 52.413 &       \\    
        83 &     KT Peg &    G2V+K5V         &  MS &      1.545 &      0.671 &      6.209 &     0.60 & 1.00 & 52.053 & 0.001 \\
        84 &     UX Ari & A2/3V+K1/2V        &  SG &      2.050 &      0.864 &      6.442 &     5.78 & 1.11 & 52.278 & 0.002 \\
        85 &     II Peg & K2IV+M0/3V         &  SG &      1.200 &      0.500 &      6.730 &          &      & 51.848 &       \\
        86 &     LR Hya & K0/1V+K1/2V        &  MS &      1.090 &      0.997 &      6.871 &     0.80 & 0.80 & 51.833 & 0.001 \\
        87 &     OU Gem &    K2V+K5V         &  MS &      1.180 &      0.831 &      6.998 &          &      & 51.889 &       \\
        88 &     SS Boo &   G0V+K0IV         &  SG &      1.928 &      0.988 &      7.603 &     3.30 & 1.30 & 52.260 & 0.001 \\
        89 &     MM Her &   G2IV+K1V         &  SG &      2.469 &      0.944 &      7.962 &     2.89 & 1.56 & 52.445 & 0.001 \\
        90 &     FF Aqr & G8III-IV+sdOB      &   G &      3.100 &      0.240 &      9.205 &     6.00 & 0.15 & 52.427 & 0.006 \\
        91 &     RU Cnc &  G8IV+F6-7         &  SG &      2.930 &      0.993 &     10.163 &     4.90 & 1.90 & 52.605 & 0.002 \\
        92 &     CQ Aur &   G8IV+F5V         &   G &      3.764 &      0.882 &     10.617 &     9.91 & 1.93 & 52.791 & 0.003 \\
        93 &     RV Lib &  G8IV-K3IV         &  SG &      2.785 &      0.183 &     10.715 &          &      & 52.294 &       \\
        94 &     EZ Peg &  G5IV+K0IV         &  SG &      1.860 &      0.991 &     11.668 &          &      & 52.296 &       \\
        95 &     42 Cap &   G2IV+G2V         &  SG &      2.375 &      0.727 &     13.183 &          &      & 52.480 &       \\
        96 &     AR Psc &   K1IV+G7V         &  SG &      2.036 &      0.818 &     14.289 &     1.50 & 1.50 & 52.387 & 0.000 \\
        97 &     TZ Tri &   K0III+F5         &   G &      5.099 &      0.976 &     14.723 &          &      & 53.060 &       \\
        98 &   V350 Lac &   K2IV-III         &   G &      1.800 &      0.818 &     17.742 &          &      & 52.329 &       \\
        99 &   zeta And &      K1III         &   G &      3.480 &      0.289 &     17.783 &    13.40 & 0.70 & 52.653 & 0.006 \\
       100 &     UV CrB &      K2III         &   G &      1.320 &      0.362 &     18.664 &          &      & 52.008 &       \\
       101 &     BL CVn &   K1II+FIV         &   G &      2.672 &      0.991 &     18.707 &    15.20 & 3.00 & 52.627 & 0.003 \\
       102 &     AR Mon & K2III+G8III        &   G &      3.413 &      0.306 &     21.232 &    10.80 & 14.20& 52.678 & 0.005 \\
       103 &     FG UMa &      G9III         &   G &      2.080 &      0.387 &     21.380 &          &      & 52.370 &       \\
       104 &     IS Vir &      K2III         &   G &      2.160 &      0.440 &     23.659 &          &      & 52.436 &       \\
       105 &     XX Tri &      K0III         &   G &      2.200 &      0.222 &     23.988 &          &      & 52.297 &       \\
       106 &     AI Phe &   F7V+K0IV         &  SG &      2.440 &      0.968 &     24.604 &     2.93 & 1.82 & 52.601 & 0.000 \\
       107 &     IM Peg &   K2III-II         &   G &      2.300 &      0.533 &     24.660 &          &      & 52.516 &       \\
       108 &     CS Cet & (G8-K1)III/IV+F    &   G &      2.780 &      0.878 &     27.353 &          &      & 52.709 &       \\
       109 &   V792 Her &  F3V+K0III         &   G &      2.881 &      0.960 &     27.542 &    12.80 & 2.58 & 52.737 & 0.002 \\
       110 &     TW Lep & F6IV+K2III         &   G &      1.990 &      0.951 &     28.314 &          &      & 52.473 &       \\
       111 &  V1762 Cyg & K2IV-III+G8V       &   G &      2.810 &      0.892 &     28.576 &     6.20 & 0.90 & 52.723 & 0.001 \\
       112 &   V965 Sco & F2IV+K1III         &   G &      3.418 &      0.990 &     30.974 &    14.00 & 5.50 & 52.878 & 0.001 \\
       113 &     RZ Eri & F0IV+G5-8III       &   G &      3.323 &      0.963 &     39.265 &     6.94 & 2.84 & 52.892 & 0.000 \\
       114 &  V4200 Ser & K2-3V+K2-3V        &  MS &      1.700 &      0.998 &     46.774 &     0.80 & 0.80 & 52.432 & 0.000 \\
       115 &     EL Eri &   G8III-IV         &   G &      1.930 &      0.379 &     48.306 &          &      & 52.430 &       \\
       116 &     AY Cet &  WD+G5IIIe         &   G &      2.640 &      0.263 &     56.885 &    15.00 & 0.012& 52.598 & 0.002 \\
       117 &     DK Dra & K1III+K1III        &   G &      3.480 &      0.981 &     64.417 &    14.00 & 14.00& 52.997 & 0.001 \\
       118 &  V1197 Ori &      K4III         &   G &      1.611 &      0.789 &    142.889 &          &      & 52.549 &       \\
       119 & BD +44 801 &  G2III+F2V         &   G &      2.780 &      0.853 &    963.829 &          &      & 53.223 &       \\
\hline
\end{tabular}  
}
\end{table*}

\begin{table*}
\caption{Physical parameters of the W UMas. First 26 are possible members of young MG ($<$ 0.6 Gyr) and the rest are older (5.47 Gyr) field systems.}
\center
{\scriptsize
\begin{tabular}{cllcccccccc}
\hline
        ID &  Star name &   Spectrel &       Type & $M_{tot}$   &       $q$  &      $P$   &   $R_{1}$ & $R_{2}$ &$\log J_{o}$ & $J_{spin}/J_{o}$\\
           &            &     Type   &            &($M_{\odot}$)&            &    (days)  &   ($R_{\odot}$) &($R_{\odot}$)& (cgs)   & \\ 
\hline
         1 &     44 Boo &        K2V &         W  &      1.339 &      0.487 &     0.2678 &     0.795 & 0.601 & 51.457 & 0.055 \\
         2 &     GZ And &        G5V &         W  &      1.625 &      0.514 &     0.3050 &     1.032 & 1.032 & 51.624 & 0.056 \\
         3 &     TW Cet &        G5V &         W  &      1.964 &      0.530 &     0.3117 &     1.048 & 0.819 & 51.768 & 0.044 \\
         4 &     SW Lac &     K0Vvar &         W  &      1.738 &      0.851 &     0.3207 &     0.997 & 0.925 & 51.724 & 0.047 \\
         5 &     RZ Com &        F8V &         W  &      1.592 &      0.437 &     0.3385 &     1.088 & 0.728 & 51.599 & 0.044 \\
         6 &     AC Boo &       F8Vn &         W  &      1.968 &      0.403 &     0.3524 &     1.314 & 0.572 & 51.744 & 0.046 \\
         7 &   V829 Her &        G2V &         W  &      1.795 &      0.408 &     0.3582 &     1.058 & 0.711 & 51.682 & 0.041 \\
         8 &     AH Cnc &        F5V &         W  &      1.830 &      0.419 &     0.3604 &     1.266 & 0.860 & 51.701 & 0.047 \\
         9 &     AE Phe &        F8V &         W  &      1.995 &      0.461 &     0.3624 &     1.269 & 0.843 & 51.781 & 0.042 \\
        10 &     AM Leo &        F5V &         W  &      2.009 &      0.449 &     0.3658 &     1.260 & 0.983 & 51.783 & 0.041 \\
        11 &     YY CrB &        F8V &         A  &      1.777 &      0.243 &     0.3766 &     1.426 & 0.805 & 51.565 & 0.062 \\
        12 &     TX Cnc &        F8V &         W  &      1.212 &      0.535 &     0.3829 &     1.019 & 0.793 & 51.450 & 0.037 \\
        13 &     BI CVn &        F2V &         A  &      2.325 &      0.413 &     0.3842 &     1.200 & 1.130 & 51.881 & 0.037 \\
        14 &     SS Ari &        G0V &         W  &      1.749 &      0.302 &     0.4060 &     1.351 & 0.795 & 51.618 & 0.047 \\
        15 &     AH Vir &        G8V &         W  &      1.772 &      0.303 &     0.4075 &     1.401 & 0.871 & 51.629 & 0.048 \\
        16 &     QX And &        F5V &         A  &      1.412 &      0.261 &     0.4118 &     1.632 & 0.804 & 51.430 & 0.078 \\
        17 &     UX Eri &        F9V &         A  &      1.876 &      0.373 &     0.4453 &     1.362 & 0.678 & 51.728 & 0.038 \\
        18 &     NN Vir &     F0/F1V &         A  &      2.676 &      0.491 &     0.4807 &     1.717 & 1.246 & 52.044 & 0.037 \\
        19 &     XZ Leo &        A8V &         A  &      2.551 &      0.348 &     0.4877 &     1.482 & 1.283 & 51.950 & 0.038 \\
        20 &     DN Cam &        F2V &         W  &      2.720 &      0.421 &     0.4983 &     1.775 & 1.224 & 52.036 & 0.040 \\
        21 &   V351 Peg &        A8V &         W  &      3.165 &      0.360 &     0.5933 &     1.875 & 1.192 & 52.141 & 0.037 \\
        22 &     UZ Leo &        A9V &         A  &      2.703 &      0.303 &     0.6180 &     2.024 & 0.851 & 51.995 & 0.043 \\
        23 & BD +145016 &        F0V &         A  &      1.918 &      0.253 &     0.6369 &     2.076 & 1.181 & 51.707 & 0.052 \\
        24 &   V753 Mon &        A8V &         W  &      3.296 &      0.970 &     0.6770 &     1.738 & 1.592 & 52.298 & 0.019 \\
        25 &     FN Cam &        A9V &         A  &      2.934 &      0.222 &     0.6771 &           &       & 51.988 &       \\
        26 &     II UMa &      F5III &         A  &      2.623 &      0.172 &     0.8250 &           &       & 51.861 &       \\
\\
        27 &     CC Com &        K5V &         W  &      1.050 &      0.522 &     0.2211 &     0.669 & 0.507 & 51.263 & 0.068 \\
        28 &   V523 Cas &        K4V &         W  &      0.950 &      0.501 &     0.2337 &     0.695 & 0.522 & 51.193 & 0.065 \\
        29 &     RW Com &        K0V &         W  &      1.230 &      0.337 &     0.2373 &     0.717 & 0.441 & 51.310 & 0.081 \\
        30 &     VZ Psc &        K5V &         A  &      1.510 &      0.911 &     0.2613 &     0.776 & 0.746 & 51.594 & 0.060 \\
        31 &     VW Cep &        G9V &         W  &      1.144 &      0.275 &     0.2783 &     1.209 & 0.398 & 51.234 & 0.097 \\ 
        32 &     BX Peg &       G4.5 &         W  &      1.400 &      0.373 &     0.2804 &     0.965 & 0.617 & 51.449 & 0.064 \\
        33 &     XY Leo &        K0V &         W  &      1.305 &      0.500 &     0.2841 &     0.899 & 0.654 & 51.451 & 0.053 \\
        34 &     RW Dor &        K1V &         W  &      1.070 &      0.672 &     0.2854 &     0.823 & 0.636 & 51.342 & 0.041 \\
        35 &     BW Dra &        F8V &         W  &      1.141 &      0.281 &     0.2923 &     0.963 & 0.542 & 51.244 & 0.083 \\
        36 &     OU Ser &     F9/G0V &         A  &      1.194 &      0.173 &     0.2968 &     0.918 & 0.426 & 51.145 & 0.113 \\
        37 &     TZ Boo &        G2V &         A  &      0.887 &      0.133 &     0.2972 &     1.050 & 0.318 & 50.846 & 0.167 \\
        38 &     FU Dra &        F8V &         W  &      1.461 &      0.251 &     0.3067 &     1.117 & 0.606 & 51.402 & 0.077 \\
        39 &     TY Boo &        G5V &         W  &      1.330 &      0.437 &     0.3171 &     1.072 & 0.761 & 51.459 & 0.054 \\
        40 &     YY Eri &        G5v &         W  &      2.160 &      0.403 &     0.3215 &     1.153 & 0.793 & 51.798 & 0.045 \\
        41 &     FG Hya &        G2V &         A  &      1.571 &      0.112 &     0.3278 &     1.371 & 0.670 & 51.216 & 0.144 \\
        42 &     AO Cam &        G0V &         W  &      1.680 &      0.413 &     0.3299 &     1.042 & 0.817 & 51.624 & 0.043 \\
        43 &     AB And &        G8V &         W  &      1.499 &      0.560 &     0.3319 &     1.041 & 0.755 & 51.589 & 0.042 \\
        44 &      W UMa &        F8V &         W  &      1.760 &      0.479 &     0.3336 &     1.126 & 0.813 & 51.684 & 0.048 \\
        45 &     EQ Tau &        G2V &         A  &      1.754 &      0.442 &     0.3413 &     1.139 & 0.786 & 51.672 & 0.048 \\
        46 &     VW Boo &        G5V &         W  &      1.400 &      0.428 &     0.3422 &           &       & 51.504 &       \\
        47 &   V757 Cen &        F9V &         W  &      1.690 &      0.690 &     0.3432 &     1.054 & 0.880 & 51.701 & 0.044 \\
        48 &   V508 Oph &        F9V &         A  &      1.530 &      0.515 &     0.3448 &     1.049 & 0.811 & 51.598 & 0.048 \\
        49 &   V781 Tau &        G0V &         W  &      1.738 &      0.405 &     0.3449 &     1.155 & 0.776 & 51.651 & 0.042 \\
        50 &     ET Leo &        G8V &         W  &      1.120 &      0.342 &     0.3465 &           &       & 51.301 &       \\
        51 &     BV Dra &        F7V &         W  &      1.398 &      0.402 &     0.3501 &     1.101 & 0.742 & 51.495 & 0.050 \\
        52 &     CK Boo &      F8/7V &         A  &      1.986 &      0.111 &     0.3552 &     1.097 & 0.815 & 51.394 & 0.096 \\
        53 &     QW Gem &        F8V &         W  &      1.700 &      0.334 &     0.3581 &     1.242 & 0.924 & 51.602 & 0.050 \\
        54 &     BB Peg &        F5V &         W  &      1.946 &      0.360 &     0.3615 &     1.286 & 0.833 & 51.717 & 0.046 \\
        55 &     LS Del &        G0V &         W  &      1.467 &      0.375 &     0.3638 &     1.046 & 0.800 & 51.522 & 0.047 \\
        56 &   V410 Aur &      G0/2V &         A  &      1.412 &      0.144 &     0.3663 &     1.397 & 0.605 & 51.239 & 0.121 \\
        57 &   V752 Cen &        F7V &         W  &      1.730 &      0.311 &     0.3702 &     1.280 & 0.754 & 51.604 & 0.049 \\
        58 &   V417 Aql &        G2V &         W  &      1.900 &      0.362 &     0.3703 &     1.290 & 0.825 & 51.705 & 0.045 \\
        59 &     XY Boo &        F5V &         A  &      1.081 &      0.158 &     0.3706 &     1.245 & 0.566 & 51.077 & 0.113 \\
        60 &     HV Aqr &        F5V &         A  &      1.564 &      0.145 &     0.3745 &           &       & 51.319 &       \\
        61 &      U Peg &        G2V &         W  &      1.528 &      0.330 &     0.3748 &     1.207 & 0.735 & 51.529 & 0.051 \\
        62 &     RT Lmi &        F7V &         A  &      1.774 &      0.367 &     0.3749 &     1.272 & 0.838 & 51.659 & 0.043 \\
        63 &     HX UMa &        F4V &         A  &      1.720 &      0.291 &     0.3792 &           &       & 51.588 &       \\
        64 &     EE Cet &        F2V &         W  &      1.829 &      0.315 &     0.3799 &     1.350 & 0.820 & 51.651 & 0.046 \\
        65 &     AU Ser &        K0V &         A  &      1.567 &      0.701 &     0.3865 &     1.051 & 0.950 & 51.665 & 0.039 \\
\end{tabular}  
}
\end{table*}

\begin{table*}
\contcaption{}
\center
{\scriptsize
\begin{tabular}{cllcccccccc}
\hline
        ID &  Star name &   Spectrel &       Type & $M_{tot}$   &       $q$  &      $P$   &   $R_{1}$ & $R_{2}$ & $\log J_{o}$ & $J_{spin}/J_{o}$\\
           &            &     Type   &            &($M_{\odot}$)&            &    (days)  &   ($R_{\odot}$) &($R_{\odot}$)& (cgs)   & \\ 
\hline
        66 &     BH Cas &        K4V &         W  &      1.095 &      0.474 &     0.4059 &     1.114 & 0.751 & 51.367 & 0.037 \\
        67 &     HT Vir &        F8V &         A  &      2.280 &      0.812 &     0.4077 &     1.262 & 1.252 & 51.953 & 0.034 \\
        68 &     EX Leo &        F6V &         A  &      1.866 &      0.199 &     0.4086 &           &       & 51.557 &       \\
        69 &   V839 Oph &        F7V &         A  &      2.136 &      0.305 &     0.4090 &     1.462 & 0.989 & 51.766 & 0.047 \\
        70 &   V566 Oph &        F4V &         A  &      1.826 &      0.243 &     0.4096 &     1.515 & 0.769 & 51.597 & 0.060 \\
        71 &     UV Lyn &        F6V &         W  &      1.844 &      0.367 &     0.4150 &     1.387 & 0.889 & 51.702 & 0.040 \\
        72 &     RZ Tau &        A7V &         A  &      2.516 &      0.540 &     0.4157 &     1.594 & 1.055 & 51.992 & 0.040 \\
        73 &     BL Eri &        G0V &         W  &      0.939 &      0.542 &     0.4169 &     1.021 & 0.777 & 51.279 & 0.031 \\
        74 &   V842 Her &        F9V &         W  &      1.713 &      0.260 &     0.4190 &           &       & 51.571 &       \\
        75 &      Y Sex &         F8 &         A  &      1.430 &      0.182 &     0.4198 &     1.240 & 0.620 & 51.341 & 0.073 \\
        76 &   V899 Her &        F5V &         A  &      2.875 &      0.566 &     0.4212 &     1.504 & 1.171 & 52.096 & 0.036 \\
        77 &     AK Her &        F5V &         A  &      1.610 &      0.229 &     0.4215 &     1.657 & 0.820 & 51.494 & 0.077 \\
        78 &     ER Ori &        F7V &         W  &      2.150 &      0.552 &     0.4234 &     1.388 & 1.130 & 51.883 & 0.035 \\
        79 &     EF Dra &        F9V &         A  &      2.102 &      0.160 &     0.4240 &     1.704 & 0.779 & 51.582 & 0.097 \\
        80 &     EF Boo &        F5V &         W  &      2.349 &      0.512 &     0.4295 &     1.090 & 1.460 & 51.939 & 0.027 \\
        81 &     AP Leo &        F8V &         A  &      1.916 &      0.297 &     0.4304 &     1.507 & 0.838 & 51.689 & 0.047 \\
        82 &     AW UMa &        F2V &         A  &      1.370 &      0.070 &     0.4387 &     1.692 & 0.560 & 50.988 & 0.233 \\
        83 &   V776 Cas &        F2V &         A  &      1.948 &      0.130 &     0.4404 &     1.710 & 0.710 & 51.465 & 0.113 \\
        84 &     TV Mus &        F8V &         A  &      1.473 &      0.119 &     0.4457 &     1.643 & 0.742 & 51.235 & 0.130 \\
        85 &   V502 Oph &        G0V &         W  &      1.778 &      0.335 &     0.4534 &     1.493 & 0.935 & 51.670 & 0.043 \\
        86 &     AA UMa &        F9V &         W  &      2.192 &      0.545 &     0.4680 &     1.356 & 1.101 & 51.910 & 0.029 \\
        87 &     DK Cyg &        A8V &         A  &      2.343 &      0.325 &     0.4707 &     1.789 & 0.987 & 51.868 & 0.051 \\
        88 &   V728 Her &        F3V &         W  &      1.949 &      0.178 &     0.4713 &     1.789 & 0.898 & 51.576 & 0.082 \\
        89 &     EL Aqr &        F3V &         A  &      1.880 &      0.203 &     0.4814 &     1.730 & 0.880 & 51.591 & 0.069 \\
        90 &     AH Aur &        F7V &         A  &      1.967 &      0.169 &     0.4943 &     1.856 & 0.897 & 51.573 & 0.085 \\
        91 &     OO Aql &        G2V &         A  &      1.920 &      0.846 &     0.5068 &     1.382 & 1.283 & 51.862 & 0.029 \\
        92 &   V401 Cyg &        F0V &         A  &      2.166 &      0.290 &     0.5827 &     1.950 & 1.170 & 51.816 & 0.046 \\
        93 &$\epsilon$ Cra&      F2V &         A  &      1.940 &      0.128 &     0.5914 &     2.200 & 0.788 & 51.500 & 0.109 \\
        94 &     AQ Tuc &       F2/5 &         A  &      2.620 &      0.358 &     0.5948 &     2.027 & 1.296 & 52.003 & 0.041 \\
        95 &     RR Cen &        A9V &         A  &      2.243 &      0.210 &     0.6057 &     2.188 & 0.974 & 51.762 & 0.066 \\
        96 &   V535 Ara &        A8V &         A  &      1.980 &      0.303 &     0.6293 &     1.847 & 1.121 & 51.772 & 0.038 \\
        97 &     AG Vir &        A8V &         A  &      2.120 &      0.317 &     0.6427 &           &       & 51.835 &       \\
        98 &      S Ant &        A9V &         A  &      2.700 &      0.392 &     0.6483 &     2.070 & 1.360 & 52.056 & 0.035 \\
        99 &  V1073 Cyg &        F2V &         A  &      2.110 &      0.319 &     0.7859 &     2.154 & 1.318 & 51.862 & 0.034 \\
       100 &  V2388 Oph &        F3V &         A  &      1.954 &      0.186 &     0.8023 &     2.373 & 1.368 & 51.668 & 0.060 \\
       101 &     TY Pup &        F3V &         A  &      2.920 &      0.183 &     0.8192 &     1.787 & 1.082 & 51.957 & 0.035 \\
       102 &   V921 Her &       A7IV &         A  &      1.974 &      0.226 &     0.8774 &           &       & 51.744 &       \\
\hline
\end{tabular}  
}
\end{table*}

\end{document}